\documentclass[tightenlines,preprint,eqsecnum,floats,showpacs,nofootinbib,amsmath,amssymb,aps,prd,superscriptaddress]{revtex4}

\usepackage{graphicx}
\usepackage{amsmath,amssymb}

\def\be{\begin{equation}}
\def\ee{\end{equation}}
\def\f{\frac}
\def\tf{\tfrac}
\def\k{\kappa}
\def\g{\gamma}
\def\w{\omega}
\def\m{\mathcal{M}}
\def\dm{{\partial\m}}
\def\S{\Sigma}
\def\dS{{\partial\S}}
\def\d{\mathrm{d}}
\def\oe{{}^oe}
\def\dual{{}^\star}
\def\r{\rho}
\def\tr{\mathrm{Tr}}
\def\si{{S_\infty}}

\begin{document}
%\preprint{\vbox{\baselineskip=12pt \rightline{IGC-10/05-1} }}

\title{Surface terms, Asymptotics and Thermodynamics\\ of the Holst Action}

\author{Alejandro Corichi} \email{corichi@matmor.unam.mx}
\affiliation{Instituto de Matem\'aticas, Unidad Morelia,
Universidad Nacional Aut\'onoma de M\'exico, UNAM-Campus Morelia,
A. Postal 61-3, Morelia, Michoac\'an 58090, Mexico}
\affiliation{Center for Fundamental Theory, Institute for
Gravitation and the Cosmos, Pennsylvania State University,
University Park PA 16802, USA}
\author{Edward Wilson-Ewing} \email{wilsonewing@gravity.psu.edu}
\affiliation{Center for Fundamental Theory, Institute for
Gravitation and the Cosmos, Pennsylvania State University,
University Park PA 16802, USA}

\begin{abstract}

We consider a first order formalism for general relativity derived
from the Holst action. This action is obtained from the standard
Palatini-Hilbert form by adding a topological-like term and can be taken
as the starting point for loop quantum gravity and spin foam
models. The equations of motion derived from the Holst action
are, nevertheless, the same as in the Palatini formulation. Here
we study the form of the surface terms of the action for general boundaries
as well as the symplectic current in the covariant  
formulation of the theory. Furthermore, we analyze the behavior of the surface
terms in asymptotically flat space-times. We show that
the contribution to the symplectic structure 
from the Holst term vanishes and one obtains the same
asymptotic expressions as in the Palatini action. It then follows
that the asymptotic Poincar\'e symmetries and conserved quantities such as energy, linear
momentum and relativistic angular momentum found here are
equivalent to those obtained from the standard Arnowitt, Deser and
Misner formalism. Finally, we consider the Euclidean approach to
black hole thermodynamics and show that the on-shell Holst action, when evaluated
on some static solutions containing horizons, yields the standard thermodynamical relations.

\end{abstract}

\pacs{04.20.Ha, 4.60.Pp}
% Asymptotic Structure, Loop Quantum Gravity

\maketitle

\section{Introduction}
\label{s1}

The study of first order actions for general relativity has three
main motivations. The first one is that they are compulsory  when
the theory is coupled to fermions. The second motivation is purely
classical. Even when the classical vacuum equations of motion one
obtains from these actions are the same as those derived from the
standard (second order) Einstein Hilbert action, there might be
new subtle effects that could appear and that are independent of
the equations of motion. For instance, when the gravitational
configurations one is considering have asymptotically flat
boundary conditions, a careful treatment of the asymptotic and
boundary terms are vital in order to have a consistent
formulation as well as conserved Hamiltonians at infinity. In
this regard, it has been known for a while that a consistent
treatment of these terms in the ADM framework involves the
introduction of terms that diverge, even on shell (see
\cite{marolf-mann} for a recent treatment). The standard {\it
Palatini} first order formalism, on the other hand, does not
possess those limitations and allows for a consistent finite
formulation of boundary conditions at infinity \cite{aes}.

Another important aspect that might arise when considering first
order actions as compared to the second order
counterparts pertains to the quantization of the theory. For
instance in a path integral quantization, even if the extrema
of the two actions are the same, the complete sum over histories
might yield different amplitudes given that one is summing over
distinct histories in each case. One particular example of this
effect can be readily seen in loop quantum gravity, where the
theories that arise from the canonical quantization of different
---classically equivalent--- actions are unitarily inequivalent,
even at the kinematical level \cite{immirzi}.

The first order action that one considers in loop quantum gravity, whose
equations of motion are Einstein's equations \cite{holst}, is
known as the {\it Holst action}. There are two terms in the Holst
action, the first one being the standard Palatini-Hilbert action and a
second term sometimes denoted as the `Holst term'.  It was originally shown in
\cite{holst} that this extra term has no impact on the equations
of motion and therefore it is often regarded as a topological term.
In addition to leaving the equations of motion untouched, the Holst
term vanishes on the space of solutions.  The Holst action
is particularly interesting as it is the action that is used as
the starting point for loop quantum gravity, not only for the
canonical quantization but also in the path integral formulation
known as spin foam models \cite{alrev, crbook, ttbook}.

While the quantum theory for the Holst action has been studied in some
detail, the same is not true of the classical theory. For
instance, a detailed study of the surface terms appearing in the
action principle of the theory has not been undertaken. The
only case where surface terms have been studied is the
case of isolated horizons.  This is because an isolated horizon is
an inner boundary to a space-time containing a black hole
\cite{ak}.  The boundary term corresponding to the isolated
horizon has been studied in quite some detail \cite{ack,
ghosh,enp1, enp2} since it is the starting point for the quantum
description of the horizon degrees of freedom \cite{abck,abk}. A
counting of the states that satisfy the quantum boundary
conditions then yields the entropy of the black hole within the
loop quantum gravity formalism \cite{dl, meissner, valencia}.

Another important application of a careful study of boundary terms in action principles
is to black hole thermodynamics. Gibbons and Hawking showed that one could approximate the
path integral for Euclidean quantum gravity by taking the saddle point approximation on
some static solutions and recover the expected thermodynamic relation between, say, 
entropy and area \cite{gh}. An important question is whether these relations
depend on the particular action one is using. Thus, there could be in principle
some tension between various action principles if they were to yield different thermodynamic
relations for static black holes. This is especially true in the case of the Holst action
where different values of the Immirzi parameter yield inequivalent
quantum theories \cite{immirzi} and a particular choice has to be made to
recover the Bekenstein-Hawking entropy from the exact counting of states in the quantum
theory \cite{abck}. An intriguing question is whether the semiclassical Euclidean
approach will retain some information of the Immirzi parameter or is independent of
this choice.

As mentioned above, it was already shown in \cite{holst} that
the equations of motion for the Holst action are the same as those for
the Palatini action.  The Hamiltonian formulation of the theory was
studied in some detail in \cite{holst, barros} while the Holst term
was studied on its own in \cite{lmp}.  A good introduction to the
classical theory of the Holst action can be found, for instance, in \cite{ttbook}.
While the equations of motion, the Hamiltonian formulation of the theory
and the symplectic structure on the bulk, among others, have been examined in some
detail there has not been much attention paid to general surface terms.
%, boundary terms have mostly been ignored in the Holst action and

The goal of this paper is to fill this gap. %address this issue.
In addition to presenting the appropriate surface term for the Holst action,
we obtain the simplified form of the surface term in the covariant
Hamiltonian description of the theory when we restrict the theory to solutions to
some of the equations of motion. Furthermore, we
study the asymptotics of the theory when considering asymptotically flat boundary
conditions, and find the corresponding conserved Hamiltonians. We thus show that the
Hamiltonians for energy, linear momentum and relativistic angular
momentum obtained for asymptotically flat space-times are the same
as those in the Palatini and ADM frameworks obtained in \cite{aes, adm}, respectively.
Finally, we consider
some physically relevant Euclidean `static' solutions and show that
there is no further contribution to the path integral from the Holst term, thus
the well defined thermodynamic behavior already found in \cite{ls2} is recovered.

It is interesting to point out that, although the conserved quantities at
infinity are the same as in the ADM framework, the surface terms in general
are not.  Indeed, in the ADM framework the surface term explicitly refers
to an embedding of the space-time into some ``background'' metric (when it exists)
\cite{hh, gh} whereas our surface term is well-defined without any
reference to a background.  When there is a background available, the two
descriptions match.  This is exactly what happens in the Palatini case
as well \cite{aes}, this is because the first order formalism (i.e., tetrads
and connections
as the basic variables rather than metrics) gives a surface term which
does not refer to any background.

A related work to this paper is Thiemann's study of the asymptotics
of the self-dual action of general relativity \cite{tt}.  The difference
between that work and ours is that in \cite{tt} the Barbero-Immirzi parameter
$\g$ is $\pm i$ in the self-dual action, whereas in the Holst action it can
be any real positive number.  The major additional difficulty that arises when
studying the Holst action is that the simplifications that
occur in the self-dual action where $\g^2 = -1$ are absent in the Holst action.
Since it is the Holst action which is considered to be the relevant action
for loop quantum gravity and spin foam models,
it is important to study the case of a real-valued Barbero-Immirzi
parameter.

There are some additional topological terms besides the Holst term that can
also be added to the Palatini action without changing the equations of motion, such as
the Nieh-Yan, Euler and Plebanski terms \cite{perez, mercuri, randono},
but we shall not consider them here.

The structure of the paper is as follows. In Sec.~\ref{s2} we
consider the Holst action and the boundary terms that need to be
added to make it consistent. We describe the kinematics of the covariant
description in Sec~\ref{s2.1} and present simplifications that arise
in the space of solutions in Sec.~\ref{s2.2}.  In Sec.~\ref{s3} we consider
asymptotically flat spacetimes within the covariant formalism. We
review in detail the asymptotic conditions that are imposed and
show that the action and symplectic structure are well defined,
finite and conserved. In Sec.~\ref{s4.a} we consider asymptotic
symmetries and find explicit expressions for the energy, linear
momentum and relativistic angular momentum. We show that, given that
the symplectic structure for asymptotically flat boundary conditions 
coincides with its Palatini counterpart, one recovers the standard
results found in the Palatini formalism. We then consider in Sec.~\ref{s4.b}
the Euclidean quantum gravity approach to black hole thermodynamics 
and show that the Holst term does not contribute to the calculation 
of the partition function for some static black hole space-times.
We end  with a discussion
in Sec.~\ref{s5}.

\section{The Holst Action}
\label{s2}

In this section, we introduce the Holst action and the appropriate surface
terms.  The first subsection will explore the surface terms in the covariant
formulation of the theory while in the second we will present some simplifications
which arise in the space of solutions.

\subsection{The Covariant Formalism}
\label{s2.1}

The independent variables in the Holst action are the co-tetrads $e_a^I$
and the Lorentz connections $\w_a{}^{IJ}$ on the space-time manifold $\m$.
The internal indices $I, J, K, \ldots$ are raised and lowered by the
Minkowski metric $\eta_{IJ}$ while the space-time indices $a, b, c, \ldots$ are
raised and lowered by the space-time metric $g_{ab} = e_a^I e_b^J \eta_{IJ}$.
The connection is antisymmetric with respect to its internal indices and it
defines the derivative operator $\nabla_a v_I = \partial_a v_I + \w_{aI}{}^J k_J$,
where the object $\w_a{}^{IJ}$ only acts on internal indices.  The curvature of
$\w$ is given by $F^{IJ} = \d \w^{IJ} + \w^{IK} \wedge \w_K{}^J,$ and
from the co-tetrads, we can construct the variable $\S^{IJ} = \dual(e^I \wedge e^J)
= \tf{1}{2}\epsilon^{IJ}{}_{KL} (e^J \wedge e^K)$.  The Holst action in the bulk
is given by the following combination of $\S$ and $F$ \cite{holst}:
\be S_b(\w, e) = \int_\m \mathcal{L} = - \f{1}{2\k} \int_\m \S^{IJ}
\wedge \left(F_{IJ} + \tf{1}{\g} \dual F_{IJ}\right), \ee
where $\k=8\pi G$ and $\g$ is the Barbero-Immirzi parameter.  The first term
is the Palatini action whereas the second term, the {\it Holst term}, does not
affect the equations of motion.  Because of this, it is often called a
topological term even though it is not a total divergence.

One can vary the action with respect to $\w$ and $e$ in order to obtain the
equations of motion, this gives
\be \label{eom1} \f{\delta S}{\delta \w^{IJ}} = 0 \quad \Rightarrow \quad
\d\Sigma^{IJ} = 0, \ee
and
\be \label{eom2} \f{\delta S}{\delta e^I} = 0 \quad \Rightarrow \quad
\epsilon^I{}_{JKL} e^J \wedge \left(F^{KL} + \tf{1}{\g}\dual F^{KL}
\right) = 0. \ee
The first equation of motion, Eq.~(\ref{eom1}) implies that
%the connection is compatible with the tetrad,
$\nabla_{[a} e_{b]}{}^I = 0$, or, equivalently,
\be \label{relwe} \w_a{}^{IJ} = e_b^I D_a e^{bJ}, \ee
where the derivative operator $D_a$ ignores internal indices.  This
also imposes that the curvature of the connection is related to the Riemann tensor
by $R_{abcd} = F_{ab}{}^{IJ} e_{cI} e_{dJ}$.  It is then straightforward to see
(see, e.g., \cite{ttbook, holst}) that the second equation of motion,
Eq.~(\ref{eom2}), implies that
\be \label{eom3} \epsilon^I{}_{JKL} e^J \wedge F^{KL} = 0, \qquad \mathrm{and}
\qquad \epsilon^I{}_{JKL} e^J \wedge \dual F^{KL} = 0. \ee
Combining these equations, one recovers Einstein's equations in vacuum%
\footnote{If matter is added to the system the equations of motion become, as they must,
$R_{ab}-\tf{1}{2}R g_{ab} = 8 \pi G T_{ab}.$}.
One can check that there are no constraints beyond those of the Palatini action and it
follows that the Holst action gives the same equations of motion as the Einstein-Hilbert
action does.

So far we have ignored the surface terms in the action.  It has been suggested that, in
generic first order actions (i.e., those that depend on the co-tetrad and the connection
rather than on the metric), the surface term should be \cite{ls}
\be S_s(\w, e) = - \int_\dm \f{\delta \mathcal{L}}{\delta F^{IJ}} \wedge \w^{IJ}. \ee
We will later see that this term will ensure that the symplectic potential is finite
in asymptotically flat space-times.  If this surface term is not added to the
action in the bulk, the symplectic potential will diverge, even in asymptotically flat
space-times, and then the symplectic structure will not be well defined.

Adding the appropriate surface term, the Holst action is given by
\be S = -\f{1}{2\k} \int_\m \S^{IJ} \wedge \left(F_{IJ} + \tf{1}{\g}\dual F_{IJ}\right)
+\f{1}{2\k}\int_\dm \S^{IJ} \wedge \left(\w_{IJ} + \tf{1}{\g}\dual \w_{IJ} \right). \ee
The surface term here is exactly analogous to the surface term in the Palatini action
studied in, e.g., \cite{aes, aabook}.

To obtain the symplectic current $J$, one varies the action and writes the result so
that only the fundamental fields $(e, \w)$ are varied, not their derivatives.  For the
Holst action, this gives
\begin{align} \delta S = -\f{1}{2\k}&\int_\m \bigg[\left( F_{IJ} + \tf{1}{\g}\dual F_{IJ}
\right) \wedge \delta \S^{IJ} + \d \S^{IJ} \wedge \left( \delta \w_{IJ} + \tf{1}{\g} \dual
(\delta \w_{IJ}) \right) \nonumber \\ &\quad + \S^{IJ} \wedge \left( \delta \w_{IK} \wedge
\w^K{}_J + \w_{IK} \wedge \delta \w^K{}_J + \tfrac{1}{\g} \dual (\delta \w_{IK} \wedge
\w^K{}_J + \w_{IK} \wedge \delta \w^K{}_J)\right) \bigg] \nonumber \\ & + \f{1}{2\k} \int_\dm
\bigg[ \delta \S^{IJ} \wedge \left( \w_{IJ} + \tf{1}{\g} \dual \w_{IJ} \right) \bigg].
\end{align}
Imposing that $\delta S$ must vanish in the bulk with respect to variations in both $e$
and in $\w$ gives the equations of motion (this is what was done in Eqs.~(\ref{eom1}) and
(\ref{eom2})), whereas the boundary term is the symplectic potential 
$\theta(\delta)= (2\k)^{-1} \delta \S^{IJ} \wedge \big( \w_{IJ}
+ \tf{1}{\g} \dual \w_{IJ} \big)$.
The symplectic current is given by the exterior derivative of $\theta$,
\be J(\delta_1, \delta_2) = \delta_1 \theta(\delta_2) - \delta_2 \theta(\delta_1), \ee
where we have assumed that the two variations $\delta_1$ and $\delta_2$ commute.  It
follows that the symplectic current 3-form is
\be \label{j-gen} J(\delta_1, \delta_2) = -\f{1}{2\k} \left[ \delta_1 \S^{IJ} \wedge
\delta_2 \left( \w_{IJ} + \tf{1}{\g} \dual \w_{IJ} \right) - \delta_2 \S^{IJ} \wedge
\delta_1 \left( \w_{IJ} + \tf{1}{\g} \dual \w_{IJ} \right) \right], \ee
note that the symplectic current is closed since it is the exterior derivative of the
symplectic potential and therefore $\d J = 0$.

\subsection{Simplifications on the Space of Solutions}
\label{s2.2}

In the previous subsection, we covered the kinematics of the Holst action but many of
these expressions are simplified when they are on half-shell, that is that the connection
satisfies Eq.~(\ref{relwe}).  One can see that by imposing Eq.~(\ref{relwe}) on the
surface term in the Holst action, the action simplifies to
\be S = -\f{1}{2\k} \int_\m \S^{IJ} \wedge \left(F_{IJ} + \tf{1}{\g}\dual F_{IJ}\right)
+\f{1}{2\k}\int_\dm \left( \S^{IJ} \wedge \w_{IJ} - \tf{1}{\g} \: e^I \wedge \d e_I \right). \ee
The symplectic potential can also be simplified, one finds that
\be \theta(\delta)= \f{1}{2\k} \: \delta \S^{IJ} \wedge \w_{IJ}
+ \f{1}{\k\g} \: \delta e^I \wedge \d e_I, \ee
and this in turn simplifies the form of the symplectic current to
\be \label{j-simp} J(\delta_1, \delta_2) = -\f{1}{2\k} \left[ \delta_1 \S^{IJ}
\wedge \delta_2 \w_{IJ} - \delta_2 \S^{IJ} \wedge \delta_1 \w_{IJ} \right]
+ \f{1}{\k\g} \d\left(\delta_1 e^I \wedge \delta_2 e_I\right). \ee
Note that the contribution due to the Holst term is of the opposite sign
than in \cite{lmp} and also that the symplectic potentials do not agree.  This
difference is due to the presence of the surface term in the action.

Depending on the situation it may be more convenient to use the kinematic
relations provided in the previous subsection or the half-on shell ones given
here, both will be useful for this work.

\section{Asymptotically Flat Space-times}
\label{s3}

The results in the previous section have been obtained without specifying any
boundary conditions and so the results obtained up to this point are quite
generic.  However, for the remainder of this paper we will only consider
space-times that are asymptotically flat.  In this section, we first describe
the boundary conditions at infinity and we then derive the pre-symplectic
structure.

%It is worth noting that the Palatini action is also a first order action
%formulation of general relativity and that therefore the procedure followed
%here will be quite similar to what was done in \cite{aes}, but we will explain
%each step taken here once again for the sake of completeness.

\subsection{Asymptotic Boundary Conditions}

We are interested in space-times where the space-time metric tends to the
Minkowski metric at infinity.  To make this more precise, we choose a point
$p$ in the interior and the radial coordinate $\r$ is defined by $\r^2 =
\eta_{ab}x^ax^b$, where $x^a$ are the Cartesian coordinates of the Minkowski
metric $\eta$ with origin $p$.  For the remainder of this section, we have this
point $p$ as the origin and we will evaluate all space-like integrals either on
the slice $M_o$ with $p$ as the origin or another slice related to $M_o$ by a
Lorentz transformation and/or a finite translation.

We can now consider functions which admit a power series expansion
\be f(\r, \Phi) = \sum_{n=0}^m \f{{}^nf(\Phi)}{\r^n} + o(\r^{-m}), \ee
where $\Phi = (\chi, \theta, \phi)$ are the angles on a hyperboloid
defined by constant $\r$.  The remainder $o(\r^{-m})$ has the property
that $\lim_{\r\to\infty}\r^mo(\r^{-m}) = 0$.  Such a function is said to
admit an asymptotic expansion to order $m$.  Note that the limit
$\lim_{\r\to\infty}\r^{m+1}o(\r^{-m})$ is not necessarily well defined
as $o(\r^{-m})$ may contain terms of the form $\log\r/\rho^{m+1}$.
A tensor is said to admit an asymptotic expansion to order $m$ if all
of its components in the Cartesian chart $x^a$ do so.

Wih this groundwork laid, we can define an asymptotically flat space-time:
a space-time with metric $g$ is asymptotically flat if there exists a
Minkowski metric $\eta$ such that, outside a spatially compact world
tube, $g-\eta$ admits an asymptotic expansion to order 1 and
$\lim_{\r\to\infty}(g-\eta)=0$.

It is clear that such a space-time will have a metric that, outside of
a spatially compact world tube, has the form
\be \label{metric1} g_{ab}\d x^a \d x^b = \left(1+\f{2\sigma}{\r}\right)\d\r^2 +
2\r\f{\alpha_i}{\r} \d\r \d\Phi^i + \r^2\left(h_{ij} + \f{{}^1h_{ij}}{\r}\right)
\d\Phi^i \d\Phi^j +o(\r^{-1}). \ee
Here $\sigma, \alpha_i$ and ${}^1h_{ij}$ only depend on the angles $\Phi^i$
and $h_{ij}$ is the metric on the unit time-like hyperboloid:
\be h_{ij} \d\Phi^i \d\Phi^j = -\d\chi^2 + \cosh(\chi)^2 \d\theta^2 +
\cosh(\chi)^2\sin(\theta)^2 d\phi^2. \ee
The form of the metric in Eq.~(\ref{metric1}) can immediately be simplified
since it has been shown that, for any space-time of this form, one can find
another Minkowski metric such that the leading order off-diagonal term
$\alpha_i$ vanishes \cite{bs}.

Before continuing, we must place restrictions on the form of $\sigma$ and
${}^1h_{ij}$ in order to avoid logarithmic translations and super-translations.
The problem is that if a metric $g$ admits an asymptotic expansion with
respect to a Minkowski metric $\eta$, it also admits an asymptotic expansion
with respect to any other Minkowski metric $\eta^\prime$ so long as
$\eta-\eta^\prime$ admits an asymptotic expansion to order 1 and
$\lim_{\r\to\infty}(\eta-\eta^\prime) = 0$.  Unfortunately, any two
Minkowski metrics $\eta$ and $\eta^\prime$ which are related by a combination
of translations, Lorentz transformations, logarithmic translations and
super-translations automatically satisfy the relation $\lim_{\r\to\infty}
(\eta-\eta^\prime) = 0$ but the Poincar\'e groups of $\eta$ and $\eta^\prime$
agree if and only if the two metrics are solely related by translations and/or
Lorentz transformations (for a more in depth discussion of this issue see,
e.g., \cite{aes}).  Therefore, we cannot allow either logarithmic translations
or super-translations if we wish to select a unique Poincar\'e group at
asymptotic infinity.

To avoid this problem, we will follow \cite{aes} and first demand that $\sigma$
be symmetric about the hyperboloid,
\be \label{sigma} \sigma(-\chi, \pi-\theta, \phi+\pi) = \sigma(\chi,
\theta, \phi), \ee
this condition removes the freedom to perform logarithmic translations,
and second that
\be {}^1h_{ij} = -2\sigma h_{ij}, \ee
which removes the freedom to perform super-translations.  This is somewhat
restrictive since it is only in space-times where the leading order term
in the Weyl curvature is both purely electric and reflexion symmetric that the
asymptotic expansion of the metric can be written in such a manner.  However,
there is a large class of space-times which are of this form and, to the best of
the authors' knowledge, such conditions are necessary in order to proceed with
the analysis of the asymptotics of the theory.  These restrictions indicate
that, as wished, the only remaining freedom is to perform asymptotic
translations and Lorentz transformations on the space-time.

The asymptotic expansion of the metric is now of the form
\be \label{metric} \d s^2 = \left(1+\f{2\sigma}{\r}\right)\d\r^2 + \left(1-
\f{2\sigma}{\r}\right) \r^2 h_{ij} \d\Phi^i \d\Phi^j + o(\r^{-1}), \ee
where $\sigma$ is reflexion symmetric as described above.

It is easy to read off the asymptotic expansion of the co-tetrads, it is simply
\be e_a^I = \oe_a^I + \f{{}^1e_a^I}{\r} + \f{{}^2e_a^I}{\r^2} + o(\r^{-2}), \ee
where $\oe_a^I$ is the Minkowski space tetrad in hyperbolic coordinates
and the subleading term is
\be \label{e1} {}^1e_a^I = \sigma(2\r_a\r^I - \oe_a^I), \ee
where
\be \r_a = \partial_a \r \quad \mathrm{and} \quad \r^I = \oe^{aI} \r_a. \ee
The connection can also be expanded in a similar fashion,
\be \w_a{}^{IJ} = {}^o\w_a{}^{IJ} + \f{{}^1\w_a{}^{IJ}}{\r} + \f{{}^2\w_a{}^{IJ}}{\r^2}
+ \f{{}^3\w_a{}^{IJ}}{\r^3} + o(\r^{-3}), \ee
and then demanding that it be compatible with the tetrad, one finds that
${}^o\w_a{}^{IJ} = {}^1\w_a{}^{IJ} = 0$ and
\be \label{w2} {}^2\w_a^{IJ} = 2\r \left(2\r_a\r^{[I}\partial^{J]}\sigma - \oe_a^{[I}
\partial^{J]}\sigma - \tf{1}{\r}\oe_a^{[I}\r^{J]}\sigma \right). \ee
Note that although $\r$ appears explicitly in the relation above, the equation
is in fact independent of $\r$ since the derivatives are inversely proportional to
$\r$: $\partial_a \sigma \propto \r^{-1} \times \partial \sigma/\partial \Phi^i$
--- recall that $\partial_\r \sigma = 0$.

Without imposing further boundary conditions, we cannot specify what the form
of ${}^2e_a^I$ is and therefore we cannot restrict the form of ${}^3\w_a{}^{IJ}$
either.

\subsection{The Holst Pre-symplectic Form}

Using the symplectic current obtained in Section II, we can obtain
a pre-symplectic form if the integral $\int_\S J$ is independent of
the Cauchy surface $\S$.  One can then obtain the symplectic form
by taking the quotient by the gauge transformations, and taking
the `projection' of the pre-symplectic form. However this last step will not
be necessary for our purposes.

To show that $\int_\S J$ is independent of the Cauchy surface, we will choose some
4-manifold $\m$ bounded by two Cauchy slices $\S_1$ and $\S_2$.  We will first
consider a region $\tilde{\m}$ which is bounded by compact portions of $\S_1$ and
$\S_2$, denoted by $\tilde{\S}_1$ and $\tilde{\S}_2$, as well as a time-like
cylinder $\tau$ which joins $\partial\tilde{\S}_1$ and $\partial\tilde{\S}_2$ and
is orthogonal to $\r^a$.  Since $J$ is closed, $\d J=0$ and
\be \label{symcur} \int_{\tilde{\S}_2} J - \int_{\tilde{\S}_1} J + \int_\tau J = 0. \ee
We now want to take the limit as $\tau$ goes to the cylinder at infinity
$\tau_\infty$; in this limit we also have $\tilde{\S}_1$ and $\tilde{\S}_2$ tending
to $\S_1$ and $\S_2$.

We will first consider the integral over $\tau$, in the limit the integrand goes as
\begin{align} \lim_{\tau\to\tau_\infty}J_{abc}\epsilon^{abc} &= \lim_{\tau\to
\tau_\infty}\tr \left(\delta_{[1}\f{{}^1\S_{ab}}{\r}\right)\wedge\left(
\delta_{2]}\f{{}^2\w_c + \tf{1}{\g}\dual({}^2\w_c)}{\r^2}\right) \epsilon^{abc}
\nonumber \\ & = \lim_{\tau\to\tau_\infty} \epsilon_{IJKL}\oe_a^K(\delta_{[1}{}^1
e_b^L)\left[\delta_{2]}\left({}^2\w_c{}^{IJ} + \tf{1}{\g}\dual({}^2\w_c{}^{IJ})
\right)\right] \r^{-3}\epsilon^{abc}. \end{align}
Since the volume element goes as $\r^3$, it is clear that the above expression is
finite.  However, we must verify that it is zero; if it is not zero, this would
indicate that some of the symplectic current is leaking out at spatial infinity.
It has already been shown in \cite{aes} that the Palatini part of this integrand
is zero, this can be seen by using Eqs.~(\ref{e1}) and (\ref{w2}) in order to
expand the above equation and then one can see that each term contains either
$\r_a\epsilon^{abc}$, $\partial_\r\sigma$ or $\delta_{[1}\sigma\delta_{2]}
\sigma$, all of which are zero.

In order to verify that the Holst part of the integrand disappears as well, we
will expand the relevant terms using Eqs.~(\ref{e1}) and (\ref{w2}) again, this
gives
\be \tf{1}{\g\r^3} \epsilon_{IJKL}\oe_a^K(2\r_b\r^L-\oe_b^L)\epsilon^{IJ}{}_{MN}\left[
(2\r_c\r^M-\oe_c^M)\delta_{[1}\sigma\delta_{2]}(\partial^N\sigma)-\oe_c^M\r^N
\delta_{[1}\sigma\delta_{2]}\sigma\right]\epsilon^{abc}. \ee
It is clear that the term containing $\delta_{[1}\sigma\delta_{2]}\sigma$ vanishes
due to the antisymmetrization and, since $\r^a$ is orthogonal to $\tau$, all terms
containing $\r_a\epsilon^{abc}$ vanish as well.  The only surviving terms, after
contracting $\epsilon_{IJKL}\epsilon^{IJ}{}_{MN}$, are
\be \tf{1}{\g\r^3} \oe_a^K\oe_b^L(\delta_{[1}\sigma)\delta_{2]}\left[\oe_{cK}
(\partial_L\sigma) - \oe_{cL}(\partial_K\sigma)\right]\epsilon^{abc}. \ee
It is clear that both of these terms vanish as they contain $\eta_{ab}\epsilon^{abc}$.
It follows that $\int_{\tau_\infty}J = 0$ and this shows that there is no symplectic
current escaping at spatial infinity.

We will now study the integral of the symplectic current over the Cauchy slice $\S_1$.
We do not expect this to vanish, but we must check that it is finite.  The integrand,
as spatial infinity is approached, is given by
\be \lim_{\tilde{\S}_1 \to \S_1}J_{abc}\epsilon^{abc} = \lim_{\tilde{\S}_1 \to \S_1}
\epsilon_{IJKL}\oe_a^K(\delta_{[1}{}^1 e_b^L)\left[\delta_{2]}\left({}^2\w_c{}^{IJ} +
\tf{1}{\g}\dual({}^2\w_c{}^{IJ})\right)\right] \r^{-3}\epsilon^{abc}. \ee
Since the volume element in this case goes as $\r^2$ and we are integrating over
$\r$, this integral could contain a logarithmic divergence; the leading order term
must disappear for the potentially logarithmic divergent term to vanish.  It has
already been shown that the potentially divergent terms due to the Palatini action
vanish \cite{aes}, we will now study the terms due to Holst's modification.  As
before, Eqs.~(\ref{e1}) and (\ref{w2}) are used and then one finds that some of the
terms disappear automatically since they contain $\delta_{[1}\sigma\delta_{2]}
\sigma$, whereas the other terms, after the two $\epsilon_{IJKL}$ tensors have
been contracted appropriately, contain either $\eta_{ab}\epsilon^{abc}$ or $\r_a
\r_b\epsilon^{abc}$ and hence vanish as well.  Since the leading order term vanishes
and the subleading terms fall off at least as quickly as $\r^{-4}$, it follows
that $\int_{\S_1}J$ is finite.

One can follow exactly the same steps to show that the integral $\int_{\S_2}J$ is
finite as well and then it follows from Eq.~(\ref{symcur}) that
\be \int_{\S_1}J \: = \int_{\S_2}J. \ee
Since this relation holds for all $\m$ and hence for all $\S_1$ and $\S_2$, this
shows that $\int_\S J$ is finite and independent of the Cauchy slice $\S$.  This
means that the pre-symplectic form $\Omega(\delta_1, \delta_2)$ is well defined
and is given by (see Eqs.~(\ref{j-gen}) and (\ref{j-simp}))
\begin{align} \Omega(\delta_1, \delta_2) &= \int_\S J(\delta_1, \delta_2)
\nonumber \\ &= -\f{1}{2\k}\int_\S\left[\delta_1\S^{IJ} \wedge \delta_2 \left(
\w_{IJ} + \tf{1}{\g}\dual \w_{IJ}\right) - \delta_2\S^{IJ} \wedge \delta_1 \left(
\w_{IJ} + \tf{1}{\g}\dual \w_{IJ}\right)\right]
\nonumber \\ &= -\f{1}{2\k}\int_\S\left[\delta_1\S^{IJ} \wedge \delta_2 \w_{IJ}
- \delta_2\S^{IJ} \wedge \delta_1 \w_{IJ} \right] + \f{1}{\k\g}\int_\dS \delta_1
e^I \wedge \delta_2 e_I. \end{align}
Note that the first part of the pre-symplectic structure is the usual Palatini
term while the Holst part of the action contributes a surface term.  One can
obtain the symplectic form from $\Omega$ by quotienting out gauge transformations,
but this will not be necessary for our purposes.

We can simplify the expression above even further for asymptotically flat space-times
by studying the surface term more carefully.  Since $\delta e^I \sim \rho^{-1}$ and
area element goes as $\rho^2$, it is clear that this term is finite and that only the
leading order term contributes.  Dropping the variations in $e^I$ which are not
tangential to the 2-sphere (in this case, proportional to $\rho_a$), one finds that
the surface term is given by
\be \int_\dS \f{1}{\rho^2}\, \delta_1 \sigma \delta_2 \sigma\; \eta_{[ab]}, \ee
which is clearly zero.  Therefore, on the space of solutions for asymptotically flat
space-times, the pre-symplectic form is given by
\be \Omega(\delta_1, \delta_2) = -\f{1}{2\k}\int_\S\left[\delta_1\S^{IJ} \wedge
\delta_2 \w_{IJ} - \delta_2\S^{IJ} \wedge \delta_1 \w_{IJ} \right]\, ,\label{new-symp-st} \ee
which is the same pre-symplectic form as for the Palatini action.
To summarize, we have seen that even when the symplectic structure for the
Holst action differs from the Palatini one by a boundary term, in the case of
asymptotically flat boundary conditions, this term vanishes and both theories
have the same symplectic structure. In the next Section we shall see two 
`applications' of our treatment of the action and boundary terms for the Holst
action.

\section{Hamiltonians and Black Hole Thermodynamics}
\label{s4}

This section has two parts. In the first one, we study asymptotically 
flat space-times and consider the asymptotic Poincar\'e symmetries for the
Holst action. In the second part, we consider the Euclidean action and its role
in Euclidean quantum gravity and black hole thermodynamics.

\subsection{Asymptotic Poincar\'e Symmetries}
\label{s4.a}

If a vector $v^a$ on $M$ represents an asymptotic Poincar\'e symmetry, i.e.,
it is a Killing vector of (one of the permissible backgrounds of $g_{ab}$)
$\eta_{ab}$, then one can study the one-form in the space of solutions
\be X_v(\delta) := \Omega(\delta, \delta_v), \ee
where $\delta_v = (\mathcal{L}_v e, \mathcal{L}_v \w)$ is the corresponding
vector field on phase space induced by the spacetime vector $v^a$.  
If $X_v$ is closed, i.e., if it can be written as%
\footnote{This relationship holds up to an additive constant which is chosen
so that the Hamiltonian is zero in Minkowski space.}%
\be \delta H_v = X_v \ee
it follows that $H_v$ is a Hamiltonian which leaves the
pre-symplectic structure invariant,
$\mathcal{L}_{\delta_v}\Omega=0$.  Such a Hamiltonian will be a
conserved quantity in the space-time. Some common examples are the energy,
linear momentum and angular momentum Hamiltonians.

To derive the form of $X_v$, one must use the equations of motion, Eqs.
(\ref{eom1}) and (\ref{eom3}) as well as the linearized
field equations for $\delta$, i.e.,
\be \delta (D\S^{IJ}) = 0, \qquad \delta (\epsilon^I{}_{JKL} e^J \wedge
F^{KL}) = 0, \quad \mathrm{and} \quad \delta (\epsilon^I{}_{JKL} e^J
\wedge \dual F^{KL}) = 0. \ee
It is also useful to use the Cartan identities,
\be \mathcal{L}_v\w = v\cdot F + D(v\cdot \w) \quad \mathrm{and} \quad
\mathcal{L}_v\S = v\cdot D\S + D(v\cdot\S) - [(v\cdot \w), \S], \ee
where the internal indices have been suppressed.

By using these equations, the general expression 
\be \label{xv} X_v(\delta) := \Omega(\delta, \delta_v) =
-\f{1}{2\k}\int_\si \tr [(v\cdot (\w+\tf{1}{\g}\dual \w))\delta\S
- (v\cdot\S)\wedge \delta(\w + \tf{1}{\g}\dual \w)], \ee
becomes, due to (\ref{new-symp-st}),
\be
X_v(\delta) := \Omega(\delta, \delta_v) =
-\f{1}{2\k}\int_\si \tr [(v\cdot \w)\delta\S
- (v\cdot\S)\wedge \delta\w ], 
\ee
where the trace is defined as $\tr(AB) = A^I{}_JB^J{}_I$ and $\si$ is
the 2-sphere at spatial infinity.  That is, we recover the same expression
for the conserved quantities as in the Palatini formalism \cite{aes}. Note that
the absence of a volume term in
$X_v$ reflects the fact that general relativity is diffeomorphism invariant.

%\subsection{Energy-Momentum}
\vskip0.5cm
\noindent
{\it Energy-Momentum.} The case of the energy-momentum Hamiltonian is obtained by considering
infinitesimal asymptotic translations which we will denote by $T^a$.  We are
therefore interested in $v^a = T^a$.  Since $\delta\S \propto \r^{-1}, \w
\propto \r^{-2}$ and the area element of the 2-sphere $\si$ at infinity goes
as $\r^2$, the first term in Eq. (\ref{xv}) vanishes and only
\be \label{XT1} X_T(\delta)  = \f{1}{2\k}\int_\si \tr
[(T\cdot\S)\wedge\delta(\w)] \ee
remains.  It is easy to check that this term is finite and, using Eqs. (\ref{e1})
and (\ref{w2}), one finds that
\begin{align} X_T(\delta) &= \f{2}{\k} \int_\si [(\r\cdot T)n^b \delta
(\partial_b\sigma) + (n\cdot T)\delta\sigma] \d^2\!S_o \nonumber \\
&= \delta \left( \f{2}{\k}\int_\si [(\r\cdot T)n^b
(\partial_b\sigma) + (n\cdot T)\sigma] \d^2\!S_o \right),
\end{align}
where $\d^2\!S_o$ is the area element of the unit 2-sphere.  

Note that we can pull $\delta$ out of the integral above since $\sigma$ is the
only dynamical variable.  It then follows that, since $X_T = \delta H_T$,
\be H_T = \f{2}{\k} \int_\si [(\r\cdot T)n^b (\partial_b\sigma) +
(n\cdot T) \sigma] \d^2\!S_o. \ee
To obtain the energy, we choose $T^a$ to be a unit time-translation which is
orthogonal to the Cauchy slice being considered and then, since $n\cdot T=-1$
and $\r\cdot T=0$, we find that
\be E = \f{2}{\k}\int_\si\sigma\;\d^2\!S_o. \ee
Recall $H_T$ is equal to $-E$ here, since $H_T$ is obtained from a 1-form
and therefore, in Cartesian coordinates, $H_{\vec{T}} = (-E, P_x, P_y, P_z)$.
Note that this corrects a typo in  \cite{aes}.

\vskip0.5cm
\noindent
{\it Linear Momentum.} In order to obtain the linear momentum Hamiltonian, 
one instead chooses a $T^a$
which lives on the Cauchy slice under consideration and then
\be P\cdot T = \f{2}{\k}\int_\si(\r\cdot T)(n\cdot
\partial\sigma)\, \d^2\!S_o, \ee
in this case there is no need to worry about the signs.  Note that $\r\cdot T$
are the spherical harmonics $Y_{lm}$ with $l=1$.  Because of this, it turns out
that while the energy is given by the $Y_{00}$ harmonic, the linear momentum is
encoded in the $Y_{1m}$ harmonics.

\vskip0.5cm
\noindent 
{\it Relativistic Angular Momentum.}
In the case where $v^a$ corresponds to an infinitesimal asymptotic
Lorentz symmetry $L^a$, the situation is a little more
complicated.  There are two cases, one where $L^a$ lives solely on
the Cauchy slice $M$, in which case it corresponds to a rotation,
and the case when $L\cdot\chi$ is nonzero which corresponds to a
boost. We will treat both cases at the same time even though the
majority of the literature on surface terms tends to ignore
boosts.  In both of these cases we will assume that the vector
field $L$ is tangential to the $\r=$ constant hyperboloids, i.e.,
$L\cdot\r=0$.  Since $L^a$ is a Lorentz symmetry, its asymptotic
behaviour is given by $L^a \sim \r$.

If one follows the same procedure as in \cite{aes} one then gets the desired 
result that the Hamiltonian generated by $L^a$ is given by
\be H_L = \f{1}{2\k}\int_\si \left( \tf{1}{\r}(L\cdot{}^o\S) \wedge {}^3\w\right); \ee
as was to be expected given that both theories share the same symplectic structure
for asymptotically flat boundary conditions.

\subsection{Black Hole Thermodynamics}
\label{s4.b}

It has been known for a long time that one can approach the description of
black hole thermodynamics by means of Euclidean quantum gravity \cite{gh}.
The idea is to consider the gravitational action and Euclidean histories in
order to compute a path integral of the form
\be {\cal Z}:=\int{\cal D}[\phi]\exp\left(-\tilde{I}[\phi]\right) \ee
with $\tilde{I}[\phi]$ the Euclidean classical action. Since the exact
evaluation of such a quantity is one major open problem, Gibbons and Hawking
proposed to consider a {\it stationary phase approximation} where one
replaces ${\cal Z}$ by its {\it on-shell} evaluation 
\be \tilde{\cal Z}:=\exp\left(-\tilde{I}[\phi_0]\right),\label{class-eval} \ee
where $\phi_0$ is a solution to the classical equations of motion and satisfies
$\delta\tilde{I}[\phi_0]=0$.   Of course, in order for this procedure to yield
sensible results the quantity $\tilde{I}[\phi_0]$ has to be well defined and
finite. As was emphasized in \cite{aes} and explored in \cite{ls2} a first
order action is vital for cases when the extra terms defined by Gibbons and
Hawking for the second order action can not be constructed.

In \cite{ls2} it was shown that for (Euclidean) Schwarzschild, Taub-NUT and
Taub-bolt solutions, the first order (Palatini) action is finite and the expression 
(\ref{class-eval}) yields the corresponding thermodynamical relations%
\footnote{Recall that, given ${\cal Z}$, one can recover the average energy $\langle
E\rangle$ and entropy $S$ as: $\langle E\rangle=-\partial (\ln {\cal Z})/
\partial \beta$ and $S=\beta \langle E\rangle +\ln({\cal Z})$, where $\beta=1/T$
is the `period', in Euclidean time, needed for regularity at the horizon \cite{gh}.}.

A natural question is whether the same result is obtained when one considers the Holst
action in the evaluation of the stationary-phase approximation.  Given that the two
actions differ by a term in the bulk (and a corresponding boundary term), one could
have a potential diagreement between the different descriptions.

The first observation is that the contribution from the bulk Holst term vanishes when
considering on-shell configurations \cite{holst}. Thus, one is lead to consider only
the contribution from the boundary terms.  In this case we have also two contributions
in the case of asymptotically flat spacetimes. We have a contribution at spatial
infinity and a term at the horizon $H$ that one would have to add to the Palatini 
terms at infinity (the horizon term vanishes in that case \cite{ls2}).

Let us now consider the extra Holst term that comes in the evaluation of the action,
\be \tilde{I}^{\rm \, Holst}_{\rm surface}=\frac{1}{2\kappa\gamma}\int_{\partial M}
\Sigma^{IJ}\wedge \dual\w_{IJ} \ee
which can be rewritten, by means of the classical equations of motion
(\ref{relwe}) as
\be \tilde{I}^{\rm \, Holst}_{\rm surface}=\frac{1}{2\kappa\gamma}\int_{\partial M}
e^I\wedge \d e_I\, . \label{action-surface} \ee
The task now is to evaluate this term (\ref{action-surface}) in both the horizon
and the asymptotic region.

It is straightforward to see that the term (\ref{action-surface}) vanishes exactly
when the tetrad $e^I_a$ is diagonal. Therefore, given that the tetrads of the
spacetimes considered in \cite{ls2} can all be put in a diagonal form, the term
(\ref{action-surface}) vanishes exactly, and the partition function $\tilde{\cal Z}$
takes the same form as in the Palatini case analyzed in \cite{ls2}. 
As one might have expected, the black hole thermodynamics one recovers
from the Holst action is the same as in the Palatini case.

A question that remains open is the case of stationary black holes that cannot be
put in a diagonal form. Note that, apart from the standard problem of finding a
real Lorentzian section associated to the corresponding Euclidean metric \cite{gh},
one would need to analyse the asymptotic behavior of the term (\ref{action-surface})
in detail. We shall not attempt to do that here.

Let us end this section with a remark. As we noted in the introduction, there is some
possible tension between the strong dependence of loop quantum gravity results such as
the black hole entropy calculation on the Immirzi parameter and the
standard semiclassical result of Gibbons and Hawking that is independent of it \cite{gh}.
Our result is consistent with the standard semiclassical result, as we have seen that the
(on shell) contribution to the action is independent of the Immirzi parameter. Thus,
if there is a connection between the path integral approach and the canonical quantization
for black holes, the dependence on the Immirzi parameter would have to manifest itself at
higher orders.  It would be interesting to see if such a dependence arises as one considers
`higher loops' contributions to the partition function.

\section{Discussion}
\label{s5}

First order actions for the gravitational field are important to study for several
reasons. If one wants to regard a classical action as an effective description of a
more fundamental theory regarding both tetrads $e$ and a connection $\w$ as fundamental
variables rather than the spacetime metric $g$, then the number of possible actions
compatible with diffeomorphism invariance is severely limited \cite{perez}.  In the
case of metric theories, the possibilities are infinite.  It is then important to
explore the different actions that generalize the Palatini term and study their properties.
The first and most relevant such addition is given by the Holst action, since this
is the starting point for quantization efforts in the loop quantum gravity program.
The results presented here constitute a first step in this direction. 

Let us summarize our results. We have considered the Holst action as an extension of
the first order Palatini gravitational action where a term that does not affect the
equations of motion is added. In the first part we made a careful analysis of the
action and the extra boundary term that one needs to include to have a well defined
action principle.  Next we studied the symplectic current that one obtains in the
covariant Hamiltonian formalism.  We proved that the contribution from the Holst term
can be written as a total divergence.

In order to make contact with standard results obtained for the first order Palatini
action, we considered in detail asymptotically flat boundary conditions.  We showed that
the extra `Holst' term does not spoil any of the features that make the Palatini action
so appealing, namely the fact that it is finite and does not need any extra `counterterms'.
Next, we made a careful analysis of the symplectic structure and found that the boundary
term coming from the Holst action vanishes due to the fall-off conditions imposed by
asymptotic flatness.

Two consequences follow from these results.  The first one is that the conserved
quantities at spatial infinity, such as energy and angular momentum, coincide
for both theories. The Holst term has no effect on the asymptotic Poincar\'e
symmetries.  The second consequence is that the Euclidean action that describes
the thermodynamic properties of static Schwarzschild, Taub-NUT and Taub-Bolt
black holes does not gain a contribution from the asymptotic Holst boundary
term, so we recover the same thermodynamic properties as in Einstein gravity.

In this paper, we have seen that the Holst action yields a well defined action
principle when a suitable boundary term is added and that the action remains
finite for asymptotically flat boundary conditions without the need of extra
conditions nor counterterms. As we have shown, the Holst action also preserves
important properties present in the pure Palatini term, making them physically
equivalent at the classical and semiclassical levels.

\section*{Acknowledgements:}

We would like to thank Abhay Ashtekar, Jonathan Engle, Simone Mercuri,
Alejandro Perez and especially David Sloan for helpful discussions.  
This research was supported in part by the NSF grant PHY0854743, CONACyT grant 80118,
DGAPA-UNAM grant IN103610, the George A. and Margaret M. Downsbrough
Endowment, the Eberly research funds of Penn State, Le Fonds qu\'eb\'ecois
de la recherche sur la nature et les technologies and the Edward A. and
Rosemary A. Mebus Funds.

\end{document}